\documentclass{article}

   \usepackage[preprint]{neurips_2024}

\usepackage{hyperref}
\usepackage{makecell}
\usepackage[utf8]{inputenc} 
\usepackage[T1]{fontenc}    
\usepackage{hyperref}       
\usepackage{url}            
\usepackage{booktabs}       
\usepackage{amsfonts}       
\usepackage{nicefrac}       
\usepackage{microtype}      
\usepackage{xcolor}         
\usepackage{times}
\usepackage{soul}
\usepackage{url}
\usepackage[utf8]{inputenc}
\usepackage[small]{caption}
\usepackage{amsmath}
\usepackage{amsthm}
\usepackage{amssymb}
\usepackage{mathrsfs}
\usepackage{booktabs}
\usepackage{algorithm}
\usepackage{algorithmic}
\usepackage{multirow}
\usepackage{graphicx} 
\usepackage{float} 
\usepackage{subfigure} 
\usepackage{pifont}
\usepackage{float}
\usepackage{wrapfig}
\usepackage{booktabs}
\usepackage{natbib}
\setcitestyle{numbers,square}
\title{A Cross-Field Fusion Strategy for \\Drug–Target Interaction Prediction}

\author{
  Hongzhi Zhang \\
  School of Computer Science\\
  Wuhan University\\
  \texttt{zhanghongzhi@whu.edu.cn} \\
  \And
  Xiuwen Gong \\
  University of Technology Sydney \\
  \texttt{gongxiuwen@gmail.com} \\
  \AND
  Shirui Pan \\
  School of Information and Communication Technology \\
  Griffith University \\
  \texttt{s.pan@griffith.edu.au} \\
  \And
  Jia Wu \\
  School of Computing \\
  Macquarie University \\
  \texttt{Jia.wu@mq.edu.au} \\
    \And
  Bo Du \\
  School of Computer Science\\
  Wuhan University \\
  \texttt{dubo@whu.edu.cn} \\
  \And
  Wenbin Hu \thanks{Corresponding author} \\
  School of Computer Science\\
  Wuhan University \\
  \texttt{hwb@whu.edu.cn} \\
}

\begin{document}

\maketitle

\begin{abstract}
    Drug–target interaction (DTI) prediction is a critical component of the drug discovery process. 
    In the drug development engineering field, predicting novel drug-target interactions is extremely crucial.
    However, although existing methods have achieved high accuracy levels in predicting known drugs and drug targets, they fail to utilize global protein information during DTI prediction. This leads to an inability to effectively predict interaction the interactions between novel drugs and their targets. 
    As a result, the cross-field information fusion strategy is employed to acquire local and global protein information. Thus, we propose the siamese drug–target interaction (\pmb{SiamDTI}) prediction method, which utilizes a double channel network structure for cross-field supervised learning.
    Experimental results on three benchmark datasets demonstrate that SiamDTI achieves higher accuracy levels than other state-of-the-art (SOTA) methods on novel drugs and targets.
    Additionally, SiamDTI's performance with known drugs and targets is comparable to that of SOTA approachs. The code is available at \href{https://anonymous.4open.science/r/DDDTI-434D/}{https://anonymous.4open.science/r/DDDTI-434D/}.
\end{abstract}

\section{Introduction}
Predicting drug–protein interactions is a crucial part of the drug development process, including small molecular drug and target interaction (DTI) and large molecular drug and protein interaction.
Protein–protein interaction (PPI) \cite{bryant2022improved} is one of the studies of large molecule drug interaction with protein.
The DTI and PPI belong to specific applications of protein representation learning in various fields of drug development. 
The information contained within protein representation varies in the two application fields.
First, the DTI field is dedicated to detecting the information contained in smaller, specific regions of the protein structure (i.e., local information), such as binding pockets and activesites. Second, the PPI field focuses on capturing the protein's overall structure and biological function (i.e., global information), including protein size, and tertiary structure. Moreover, local and global protein information are interdependent, affecting the protein's biological function concurrently.
Therefore, integrating protein information from different fields can generate more comprehensive protein representations from multiple perspectives, improving the prediction accuracy of drug–protein interactions.

Small molecule DTI prediction plays a crucial role in drug discovery \cite{dehghan2023tripletmultidti,chatterjee2023improving,gao2024drugclip}, as small molecule drugs have the advantages of high chemical stability and low production cost. 
When the training data includes the drug and target type present in the test set, these drugs and targets are considered "known". Conversely, they are deemed "novel" (zero-shot learning) if they do not appear in the training data. The computational biology and medicine fields are experiencing rapid growth, resulting in the continual identification and exploration of novel targets \cite{pun2023ai}. Simultaneously, pharmaceuticals occupy only a minuscule portion of the expansive molecular realm \cite{dobson2004chemical}. Hence, predicting of novel DTIs is more closely aligned with applying drug development engineering practically.

\begin{figure}
    \centering
    \includegraphics[width=1.0\linewidth]{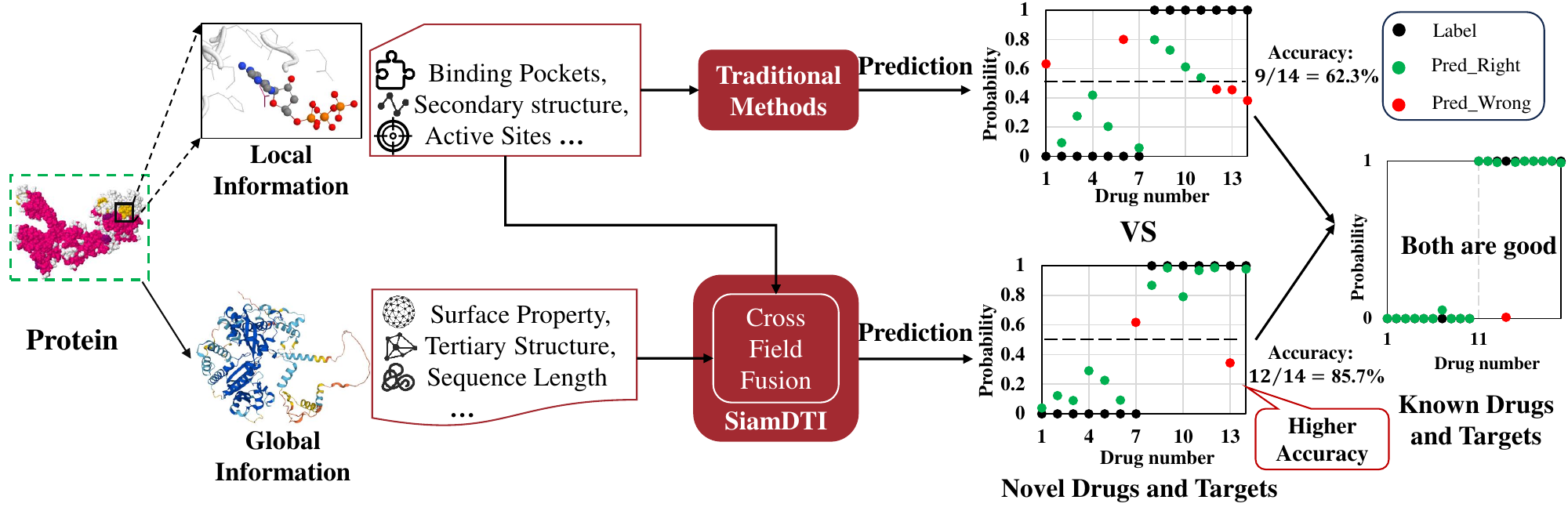}
    \caption{For novel drugs and targes, traditional methods do not utilize the global information of target proteins, resulting in poor performance. SiamDTI achieves better performance by fusing global and local protein information. For known drugs and targes, both traditional methods and SiamDTI can achieve good results.}
    \label{fig:fig1}
\end{figure}

Despite the remarkable DTI prediction progress for known drugs and targets \cite{dehghan2023tripletmultidti,tsubaki2019compound}, the current task of predicting interactions between novel ones remains highly challenging. 
For instance, the target (i.e., a protein linked to disease) encompasses highly intricate and diverse information. 
Moreover, traditional methods do not incorporate global protein information, thereby limiting the knowledge obtained from protein characterization. These challenges severely restrict the performance of existing DTI prediction methods for novel drugs and targets. For instance, in the work of \cite{nguyen2021graphdta}, the prediction performance of their models based on graph neural networks(GNNs) exceed 90\% in the area under the receiver operating characteristic curve(AUROC) for known drugs and targets. However, the AUROC decrease to a minimum of 60\% for novel drugs and targets.
Therefore, comprehensive and diverse access to the target protein's information is important. 
Figure \ref{fig:fig1} illustrates the diverse target protein information that traditional methods fail to integrate, resulting in poor performance on novel drugs and targets.

To address these issues, we employ the cross-field (i.e., PPI and DTI) information fusion strategy to integrate local and global protein information. 
We propose the siamese drug–target interaction (\pmb{SiamDTI}) prediction method, which seeks to improve prediction performance on novel drugs and targets by obtaining comprehensive protein information. 
This method incorporates the PPI and traditional DTI modules, establishing two distinct channels.
The PPI channel detects global protein information that is absent in the DTI module.
In addition, both channels share all the protein encoders, combining the cross-field protein information and enhancing representation.
Furthermore, SiamDTI can transform drug–target pairs into the novel characterization space, thereby increasing the separation between samples with different labels.
To verify the effectiveness of the SiamDTI, we conduct experiments within two different application scenarios. The results show that SiamDTI achieves higher accuracy than other DTI prediction methods for novel drugs and targets. 

\section{Related Work}

DTI studies on clinical and laboratory data has made prediction exploration feasible. 
There are two main in silico DTI prediction methods: molecular docking \cite{halgren2004glide,trott2010autodock,verdonk2003improved,pei2024fabind,jin2024unsupervised} and representation learning \cite{somnath2021multi,zhao2021identifying,gao2018interpretable,jiang2020drug}. However, the molecular docking method requires obtaining the drug molecule and target protein three-dimensional structure, has a relatively low accuracy level \cite {moon2022pignet}, and a high computational energy demand \cite{sledzieski2022adapting,sadybekov2022synthon}. This hinders the large-scale application of DTI technology.
 
Previous studies \cite{li2022predicting}\cite{shin2019self}\cite{zheng2020predicting} have introduced various representation learning-based DTI techniques that involve manually or automatically extracting drug and target protein representations. These representations are then mapped to the reaction result space to obtain the prediction results. 
The machine learning based methods aim to map manually constructed drug molecule and target protein representations to the interaction outcome space \cite{he2017simboost}. Recently, an increasing number of studies have started to leverage leveraging deep neural networks (DNNs) for extracting drug and target protein features to enhance feature representation. During this process, molecules are typically represented by simplified molecular-input line-entry system (SMILES) \cite{weininger1988smiles} sequences or molecular fingerprints \cite{rogers2010extended}, and targets portrayed as amino acid sequences \cite{lee2019deepconv}. DeepDTA \cite{ozturk2018deepdta} utilizes one-dimensional drug SMILES and target protein sequences as input. Then, it employs a one-dimensional convolutional neural network to extract molecule representations. Finally, the drug and target representations are directly concatenated and then decoded by fully connected layers to accomplish the DTI prediction. 
Similar methods also have \cite{zhang2022deepmgt}\cite{yang2021ml}\cite{mahdaddi2021ea} etc.
However, these methods ignore the molecular structure information, resulting in the model's poor performance in DTI prediction tasks.

Current research has embraced graph-based representations for drug molecules, as demonstrate in recent works like \cite{xu2021prediction} and \cite{cho2020layer}, yielding outstanding results and garnering considerable attention in the DTI prediction field.
For example, GraphDTA \cite{nguyen2021graphdta} encodes molecules as two-dimensional molecular graph structures during DTI tasks. It utilizes GNNs to obtain drug representations in the hidden space. This approach demonstrates GNNs' superior performance compared to other DNNs in predicting DTIs. Methods like \cite{wang2021drug} conduct DTI prediction by constructing topological graphs for drug and target affinity reactions.
These unimodel prediction methods perform adequately for known drugs and targets, but their generalization abilities for novel drugs and targets are insufficient.

\section{Method}
\subsection{SiamDTI Architecture}
The siamese network model is commonly employed during tasks such as object tracking \cite{guo2017learning} and signature verification \cite{bromley1993signature,chakladar2021multimodal}. 
This paper constructs a siamese network structure called SiamDTI for DTI application. SiamDTI comprises two similar channels: the DTI channel (i.e., simulating the process of drug binding to target proteins) and the PPI channel (i.e., simulating the process where identical targets do not interact). Similar to the traditional DTI prediction methods,the DTI channel consists of two branches: a drug and protein branch, while PPI channel comprises two identical drug branches. In the PPI channel, the input proteins are the same target protein, and SiamDTI treats one of it as an ineffective large molecule drug to obtain global protein information. In SiamDTI, the protein branch network structures and parameters are equal, thereby reducing the impact of model overfitting on the DTI prediction results. Both channels use the bilinear attention network to map different channel features to the same hidden space. 
The fusion feature is obtained by subtracting the features of the two channels. 
Finally, the prediction module uses a multilayer perceptron (MLP) on fusion feature to predict the interaction probability between the input drug–target pair.
\begin{figure}
    \centering
    \includegraphics[width=1.0\linewidth]{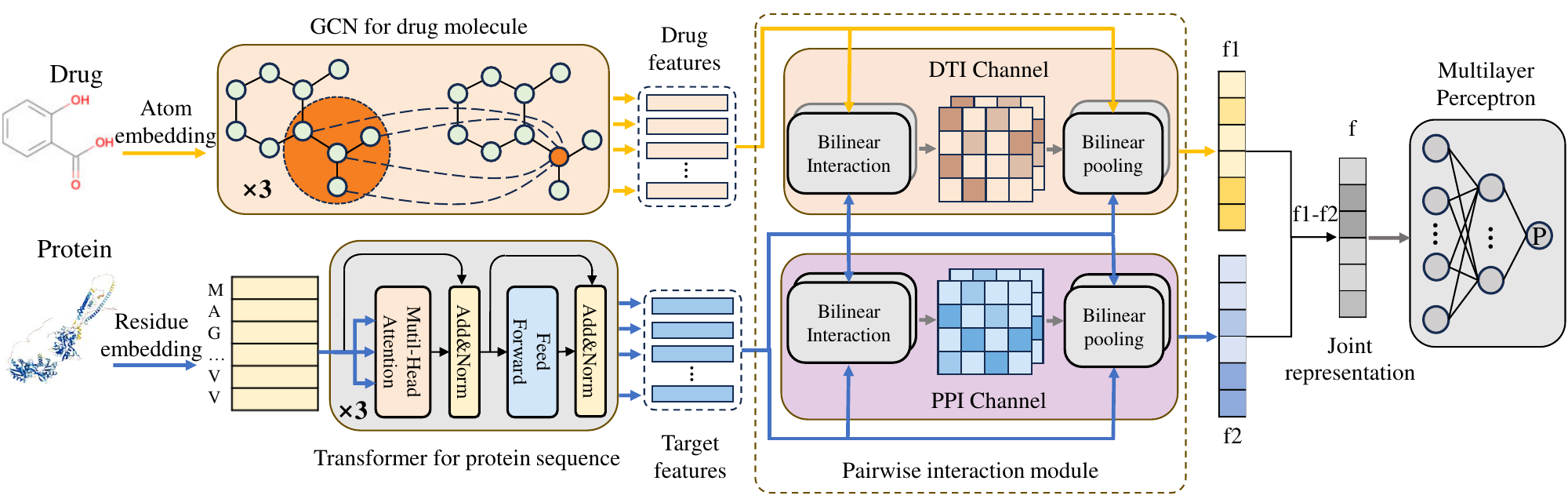}
    \caption{\textbf{Overview of the SiamDTI framework. }SiamDTI consists of two channels: drug–target interaction channel and protein–protein interaction channel. The protein input for both channels are the same target protein. It accurately predicts DTI by dual channel information fusion.}
    \label{fig:network}
\end{figure}
\subsection{Transformer for Protein Sequence}
The protein feature encoder includes a transformer encoder, which encodes the input protein amino acid sequence into a representation matrix in a latent feature space. Similar to the principles of word embeddings in natural language processing research, we  initialize the amino acid sequence as a learnable two-dimensional embedding matrix $E_p{\in}\mathbb{R}^{N_a{\times}D_p}$, which is used to map different amino acids into the real number domain. In this expression, $N_a$ represents the different amino acid types, and the unknown one are depicted as the same type. $D_p$ denotes the feature embedding's dimension. By querying the matrix $E_p$, the protein's amino acid sequence is transformed into a two-dime nsional matrix $X_p{\in}\mathbb{R}^{\Theta_p{\times}D_p}$, where $\Theta_p$ represents the protein's maximum sequence length. Furthermore, the target protein sequences are set to the same length for batch processing and model training. According to previous research, protein sequences longer than the maximum length will be truncated, and shorter ones will be padded with zeros.
Hence, the protein feature encoder extracts local residue patterns from the protein's encoded sequence. Moreover, the self-attention mechanism enables the model to obtain important information from different positions in the input amino acid sequence, thereby enhancing the model's understanding and representation of the input sequence. The protein encoding module is expressed  as follows:
\begin{equation}
H^{(l+1)}_p=\sigma(\mathrm{Transformer}(A^{(l)}_T,H^{(l)}_p)),
\end{equation}

where $A^{(l)}_T$ represents the learnable attention matrix in the transformer encoder, $H^{(l)}_p$ denotes the hidden protein encoding representation of the $l$-th layer, and $H^{(0)}_p = X_p$, $\sigma(\cdot)$ denotes the non-linear activation function (i.e., ReLU($\cdot$)).

In the self-attention mechanism, the input protein matrix $X_p$ is transformed into query, key, and value matrices, denoted as $Q, K, V \in \mathbb{R}^{N \times d}$. Also, the attention scores are normalized using the softmax formula, which is calculated according to the following formula:
\begin{equation}
\mathrm{Attention}(Q,K,V)=\mathrm{softmax}(\frac{QK^T}{\sqrt{d}})V.
\end{equation}

In this case, the self-attention mechanism receives either the input (i.e., protein representation matrix $X_p$) or output of the previous encoder block. Additionally, Q, K, V are precisely obtained through linear transformations of the input through self-attention.

\subsection{GCN for Drug Molecule}

Previous studies have demonstrated the GNN's excellent performance during DTI tasks. For example, to encode drug molecules, we transform the drug's SMILE sequence into a two-dimensional graph structure $\mathcal{G}$, where the nodes represent atoms and the edges depict chemical bonds. To represent the graph's node information, we initialize all the atoms based on their chemical properties. Each atom is represented by a 74-dimensional integer vector, describing eight different information aspects: the atom type, the atom degree, the number of implicit Hs, the formal charge, the number of radical electrons, the atom hybridization, the number of total Hs, and whether the atom is aromatic. Similar to setting the maximum protein sequence length mentioned above, we also need to manually set the maximum number of nodes $\Theta_d$. Compounds with fewer nodes than the maximum number are padded with virtual nodes filled with zeros. Therefore, each molecular graph's node feature matrix is represented as $M_d \in \mathbb{R}^{\Theta_d \times 74}$. In addition, we use a simple linear transformation $X_d = W_0 M_d^T$ to convert the input drug molecule into an input feature matrix $X_d \in \mathbb{R}^{\Theta_d \times D_d}$.

We utilize a three-layer graph convolutional network (GCN)  to process the input drug feature matrix, effectively learning drug molecule representations. GCN is employed to compute the the neighboring atoms influence on each atom and update its feature vector accordingly. The expression for drug encoding is as follows:
\begin{equation}
H^{(l+1)}_d = \sigma(\mathrm{GCN}(\tilde{A}, W^{(l)}_g, b^{(l)}_g, H^{(l)}_p)),
\end{equation}

where \( W^{(l)}_g \) and \( b^{(l)}_g \) denote the learnable weight matrix and bias vector of the \( l \)-th GCN layer, respectively. \( \tilde{A} \) represents the adjacency matrix of the molecular graph \( \mathcal{G} \). \( H^{(l)}_p \) signifies the \( l \)-th layer's hidden protein encoding representation, where \( H^{(0)}_p = X_p \).

\subsection{The Pairwise Interaction Module}
In the feature fusion module, the existing methods fail to effectively integrate protein  and drug features using linear concatenation. Additionally, if linear concatenation is employed in the fusion module, subtracting the two branches' features indicates the distance between drugs and target proteins, which does not consider the binding process between proteins and drugs, thus not suitable for the DTI prediction method proposed in this paper.
We utilized a bilinear attention network (BAN) \cite{kim2018bilinear} for feature fusion, aiming to capture reaction features between pairs of proteins or between drugs and targets. It consists of two layers: a bilinear interaction graph to capture pairwise attention weights and a pooling one to extract joint drug–target or protein–protein representations. The transformer and GCN encoders generate hidden protein and drug representations at the third layer, denoted as \( H^{(3)}_p = \{H^1_p, H^2_p, \cdots, H^M_p\} \) and \( H^{(3)}_d = \{H^1_d, H^2_d, \cdots, H^N_d\} \), where \( M \) and \( N \) represent the number of protein encoding substructures and atoms in the drug, respectively.
Furthermore, the feature matrices' fusion method is identical in both the DTI and PPI channels. Below, we describe the feature matrices fusion method in the DTI channel.

We construct a bilinear interaction graph using these hidden features to obtain a pairwise interaction matrix \( A \in \mathbb{R}^{N \times M} \):
\begin{equation}
B = ((1 \cdot q^\top) \circ \sigma((H^{(3)}_d)^\top U)) \cdot \sigma(V^\top H^{(3)}_p), 
\end{equation}

where \( U \in \mathbb{R}^{D_d \times K} \) and \( V \in \mathbb{R}^{D_p \times K} \) represent the learnable weight matrices for drug and protein representations, \( q \in \mathbb{R}^{D_p \times K} \) is a learnable weight vector, \( 1 \in \mathbb{R}^N \) is a fixed matrix of 1s, and \( \circ \) denotes the Hadamard product. The elements in matrix \( B \) indicate the interaction strength of drug–target or target–target pairs. As we used two bilinear attention networks with the same structure but different weights, allowing us to learn different interaction representations. 
The value $ {\rm I}_{ij} $ in the matrix I can be written in the following form:
\begin{equation}
B_{ij} = q^\top(\sigma(U^\top h_d^i) \circ \sigma(V^\top h_p^j)),
\end{equation}

where \( h_d^i \) represents the \( i \)-th column of \( H^{(3)}_d \), and \( h_p^j \) represents the \( j \)-th column of \( H^{(3)}_p \), denoting the drug and protein's the \( i \)-th and \( j \)-th substructures, respectively. First, \( H^{(3)}_d \) and \( H^{(3)}_p \) are mapped to the same space through the learnable matrices \( U \) and \( V \). Then the interaction is computed using the Hadamard product and the weight matrix \( q \).

To obtain the joint representation \( f' \in \mathbb{R}^K \), we applied a bilinear pooling layer on the reaction matrix \( B \). The calculation for the \( k \)th element of \( f' \) is as follows:
\begin{equation}
f'_k = \sigma\left((H^{(3)}_d)^\top U_k^\top \cdot { B} \cdot \sigma\left((H^{(3)}_p)^\top V\right)_k\right)  = \sum_{i=1}^{N} \sum_{j=1}^{M} { B}_{i,j}(h^i_d)^\top(U_kV_k^\top)h^j_p,
\end{equation}

where \( U_k \) and \( V_k \) represent weight matrices \( k \)th column. There are no learnable parameter matrices in this layer. Matrices \( U \) and \( V \) share the interaction matrix graph from previous layers to reduce the parameters and avoid overfitting. We also perform sum pooling on the joint representation matrix to obtain the DTI feature map:
\begin{equation}
f_{dt} = {\rm SumPool}(f', s),
\end{equation}
\( \mathrm{SumPool(\cdot)} \) is a one-dimensional non-overlapping stride \( s \) pooling operation. It reduces the dimensionality of \( f' \) from \( \mathbb{R}^K \) to \( \mathbb{R}^{K/s} \).
We obtain the feature map for protein–protein interactions \( f_{pp} \) by setting both input matrices of the bilinear interaction graph to \(H^{(3)}_p\).

Using the bilinear attention mechanism, the model accurately learns the interactions between drug–target and protein–protein pairs. In addition, by subtracting the two different joint representations, we obtain the difference in feature representations between channels \( f\):
\begin{equation}
f = f_{dt}-f_{pp}\label{con:F}.
\end{equation}

SiamDTI uses  \( f \)  to combine different channel features which obtain more protein information.
Due to channels' fusion feature, the SiamDTI can distinguish samples with different labels to achieve high DTI prediction accuracy. To compute the interaction probability, we input the difference in feature representations between the two channels \( f \) into a decoder consisting of a fully connected layer, and use the sigmoid function as the activation function:
\begin{equation}
p = \mathrm{Sigmoid}(W_0f + b_0)\label{con:Sigmoid},
\end{equation}
where \( W_0 \) and \( b_0 \) represent the learnable weight matrix and bias vector, respectively.

Finally, we perform backpropagation to update all the optimized learnable parameters. The loss function used for model training is the cross-entropy loss, calculated as follows:
\begin{equation}
L = -\sum_i (y_i \log(p_i)) + (1 - y_i) \log(1 - p_i) + \frac{\lambda}{2} \| \theta \|_2^2\label{con:Loss},
\end{equation}
where \( \theta \) represents all learnable weight matrices and bias vectors, \( y_i \) represents the ground truth labels for all drug-target pairs, \( p_i \) is the probability of a drug-target pair reacting as output by the model, and \( \lambda \) is a hyperparameter used for L2 regularization.

\subsection{Complexity Analysis}
\begin{table}[!htbp]
\caption{Computational analysis and performance comparison of different models. Performance measured by the number of Parameters/FLOPs.}
\setlength{\tabcolsep}{15pt}
\renewcommand\arraystretch{1.2}
    \centering
    \begin{tabular}{ccc}
        \hline
                   Method                   &  Para.(M)                 & FLOPs($10^6$)    \\
        \hline 
                   DeepConv-DTI             & $ 1.49$                   & $8.93$  \\
                   GraphDTA                 & $ 4.07$                   & $34.79$  \\
                   MolTrans                 & $ 47.16 $                 & $2268.77$  \\
                   HyperAttentionDTI        & $ 2.3 $                   & $2356.84$ \\
                   DrugBAN                  & $ 1.0$                    & $515.67$  \\
                   SiamDTI                  & $4.96$                    & $4611.06$  \\
        \hline
    \end{tabular}

    \label{tab:table3}
\end{table}

We conduct computational analyses to compare the various methods' time consumption and parameter quantities. The experiments use an Intel Xeon E5-2690 v3 12-core 24-thread processor with a clock frequency of 2.60 GHz. In addition, an RTX 4090 GPU is used.
As shown in Table \ref{tab:table3},SiamDTI uses transformer for protein coding, so its floating point operations (FLOPs) metric is the largest. However, the transformer can reduce the impact of large FLOPS due to its parallel computation capacity. In addition, the SiamDTI method's parameter quantity is about one-tenth of that of MolTrans, reflecting that SiamDTI uses fewer parameters. Notably, All our protein coding modules adopt the parameter-sharing transformer encoder module, significantly reducing the model's parameters and improving its generalization capacity.
In summary, the model is expected to integrate more information using the two-channel framework with a small increase in the parameter number.
\label{Complexity Analysis}
\section{Experiment}
\textbf{Task Definition. }Given a set of DTI sequences \(E = \{e_{ij} = \{p_i, g_j\} | p_i \in P, g_j \in G, Y(e_{ij}) \in \{0, 1\}\}\), where \(Y\) represents a binary indicator, with \(Y = 1\) indicating an interaction between the drug and the target protein, and \(Y = 0\) indicating no interaction, DTI prediction aims to train a model to map the joint feature representation space as an interaction probability score.

\textbf{Experimental Settings. }We adhere to the DrugBAN method in the DTI task network parameter settings. We train the SiamDTI model for 150 epochs with an initial learning rate of 5e-5 and a batch size of 64, using the Adam optimizer. The optimal model, which achieves the highest AUROC score on the validation set, is selected and used to evaluate the test set and assess the model's performance. Furthermore, the maximum protein sequence length is set to 1200, and the maximum atom number for the drug molecules is set to 290. Finally, the protein encoding module consisted of three identical transformer encoders, each containing eight attention heads. In SiamDTI, the three protein encoding modules' structures are identical, and their parameters are shared to reduce the parameter number and mitigate overfitting.In the protein coding module, the amino acid type $N_a$ is set to 23, where all unknown amino acid types are counted as one. In the bilinear attention module, we use two attention heads with a hidden embedding size \( k \) of 168 and a pooling window size \( s \) of 3. The hyperparameters for the bilinear attention module are equal in both channels. Finally, the hidden neurons number is set to 512 in the fully connected layer decoder, and our model's performance is not affected by the hyperparameter settings. 
\label{experiment}


\textbf{Baseline Methods. }We compare SiamDTI with several representative DTI methods, including DeepConv-DTI \cite{lee2019deepconv}, GraphDTA \cite{nguyen2021graphdta}, MolTrans \cite{huang2021moltrans}, HpyerAttentionDTI \cite{zhao2022hyperattentiondti} and DrugBAN\cite{bai2023interpretable}, which use the protein amino acid sequences as input. DeepConv-DTI, HpyerAttentionDTI, and MolTrans use the drugs' SMILE expressions as input, while GraphDTA and DrugBAN convert the drug molecules' SMILE expressions into a graph structure and use that as input. The GraphDTA's application involves predicting the drug–target binding affinity values. In the final part of GraphDTA, we add a Sigmoid activation function to convert it into a binary classification problem and adjust its hyperparameters using two-dimensional cross-entropy loss. 
Furthermore, we conduct extensive comparative experiments between SiamDTI and these benchmark methods within various scenarios. The evaluation metrics and datasets used in this article can be found in the Appendix \ref{app:Metrics} and \ref{app:Datasets}.
\begin{figure}
    \centering
    \includegraphics[width=1.0\linewidth]{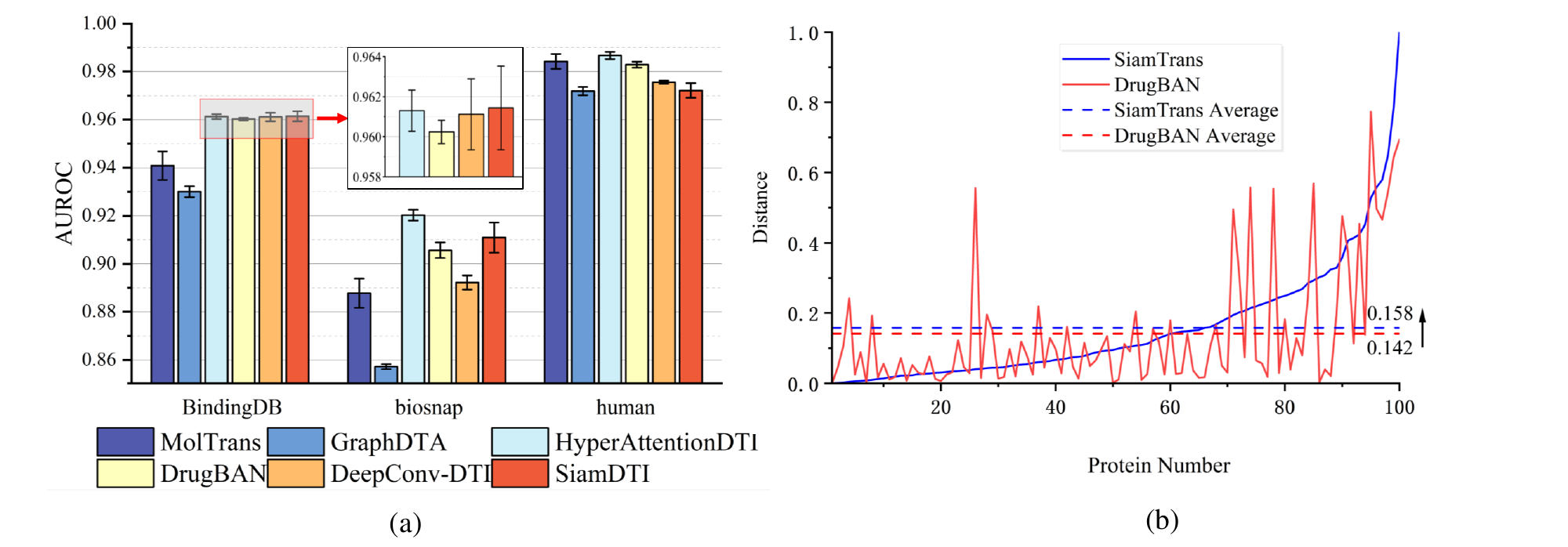}
    \caption{(a) Performance comparison of the methods for known drugs and targets on the three datasets.The histogram represents the average results of five randomized experiments, and the error bars are the size of the standard deviation. (b) Comparison of the SiamDTI and DrugBAN for sample representation distances of known drugs and targets on BindingDB\cite{gao2018interpretable}. The horizontal coordinate represents the target protein, and the vertical coordinate denotes the representation distance of different labeled drug target pairs composed of that target. A larger average distance indicates better classification performance.}
    \label{fig:fig3}
\end{figure}
\section{Results}

\subsection{The Performance Comparison on Known Drugs and Targets}
In this scenario, each experimental dataset is randomly divided into training, validation, and test sets in a 7:1:2 ratio. Based on this setup, we compare SiamDTI with other baseline methods, including DeepConv-DTI, GraphDTA, MolTrans, HyperAttentionDTI, and DrugBAN. Figure \ref{fig:fig3} presents the comparative results on the BindingDB, BioSNAP \cite{wishart2008drugbank}, and human\cite{liu2015improving} datasets. SiamDTI generates optimal and suboptimal AUROC scores for the BindingDB and BioSNAP datasets. This can be attributed to its more effective acquisition of interaction features between drugs and targets.
However, SiamDTI's prediction performance for the human dataset is inadequate. This may be due to the human dataset having a small number of samples, and the sample fluctuation leading to an approximately 1\% difference between SiamDTI and best method. Figure \ref{fig:fig3} (b) reports distances between positive and negative drug–target pair features for 100 targets on the BindingDB dataset. Figure \ref{fig:fig3} illustrates that SiamDTI can improve the distance between positive and negative sample features more than the baseline, and the average distance is increased by 11.3\%.

Overall, all the methods perform adequately within this scenario, especially on the human dataset, where all methods achieve an AUROC greater than 97\%, indicating relatively easy prediction tasks. The known drugs and targets setting is particularly valuable for drug repurposing, where large drug libraries are evaluated against known proteins implicated in a disease of interest.

\begin{table*}
     \caption{The performance comparison on the BindingDB, BioSNAP and human datasets for novel drugs and targets (statistics over five random runs). The results are presented as mean$\pm$standard deviation. The beat performance for each dataset and metric has been highlighted in \pmb{bold}.} 
     \setlength{\tabcolsep}{8pt}
    \renewcommand\arraystretch{1.1}
    \centering
    \begin{tabular}{ccccc}
        \hline
        Benchmark & Method            & AUROC             &      AUPRC        & F1    \\
        \hline
         
        \multirow{6}*{BindingDB}
                  & DeepConv-DTI      & $0.527\pm{0.038}$        & $ 0.499\pm{0.035}$        & $0.398\pm{0.135}$  \\
                  & GraphDTA          & $0.567\pm{0.029}$        & $ 0.535\pm{0.028}$        & $0.473\pm{0.012}$  \\
                  & MolTrans          & $0.588\pm{0.014}$        & $ 0.534\pm{0.015}$        & $0.300\pm{0.162}$  \\
                  & HyperAttentionDTI & $0.572\pm{0.030}$        & $ 0.520\pm{0.030}$        & $0.472\pm{0.034}$  \\
                  & DrugBAN           & $0.555\pm{0.027}$        & $ 0.524\pm{0.035}$        & $0.668\pm{0.001}$  \\
                  & SiamDTI           & $\pmb{0.627\pm{0.010}}$  & $\pmb{0.571\pm{0.024}}$   & $\pmb{0.681\pm{0.010}}$  \\
        \hline
        \multirow{6}*{BioSNAP}
                  & DeepConv-DTI      & $0.649\pm{0.017}$         & $0.653\pm{0.020}$         & $0.353\pm{0.180}$  \\
                  & GraphDTA          & $0.556\pm{0.022}$         & $0.549\pm{0.025}$         & $0.155\pm{0.104}$  \\
                  & MolTrans          & $0.621\pm{0.014}$         & $0.618\pm{0.014}$         & $0.325\pm{0.166}$  \\
                  & HyperAttentionDTI & $0.679\pm{0.012}$         & $0.693\pm{0.011}$         & $0.450\pm{0.034}$  \\
                  & DrugBAN           & $0.616\pm{0.010}$         & $0.610\pm{0.013}$         & $0.671\pm{0.003}$  \\
                  & SiamDTI           & $\pmb{0.718\pm{0.006}}$   & $\pmb{0.725\pm{0.010}}$   & $\pmb{0.697\pm{0.005}}$  \\
        \hline
        \multirow{6}*{human}
                  & DeepConv-DTI      & $0.761\pm{0.016}$         & $0.628\pm{0.022}$       & $0.627\pm{0.030}$ \\
                  & GraphDTA          & $0.822\pm{0.009}$         & $0.759\pm{0.006}$       & $0.688\pm{0.030}$ \\
                  & MolTrans          & $0.810\pm{0.021}$         & $0.745\pm{0.034}$       & $0.648\pm{0.042}$ \\
                  & HyperAttentionDTI & $0.854\pm{0.009}$         & $0.779\pm{0.015}$       & $0.636\pm{0.057}$ \\
                  & DrugBAN           & $0.833\pm{0.020}$         & $0.760\pm{0.031}$       & $0.785\pm{0.015}$ \\
                  & SiamDTI           & $\pmb{0.863\pm{0.019}}$   & $\pmb{0.807\pm{0.040}}$ & $\pmb{0.803\pm{0.016}}$ \\
        \hline
    \end{tabular}
   
    \label{tab:table1}
\end{table*}

\subsection{The Performance Comparison on Novel Drugs and Targets}
In this instance, the training and test data do not contain the same drugs and targets, making it impossible to rely on the known drug and target features when predicting test data. Nevertheless, this scenario is suitable for large-scale drug virtual screening against new targets, where almost all the molecules in the drug screening library are novel such as the molecular library generated by diffusion model. 

The experimental results are shown in the Table \ref{tab:table1}. The results indicate that, in this scenario, the performance of all methods is inferior to that in the known drugs and targets scenario due to obtaining limited information.
In comparison, the SiamDTI achieves state-of-the-art (SOTA) performance in the AUROC and area under the precision-recall curve (AUPRC) metrics. On the BindingDB dataset, the SiamDTI method improves the AUROC and AUPRC metrics by 7.6\% and 13.7\%, respectively, compared to the DrugBAN. On the BioSNAP dataset, the SiamDTI method enhances the AUROC and AUPRC metrics by 9.7\% and 15.2\%, respectively, compared to DrugBAN. Simultaneously, compared with the suboptimal method, SiamDTI improves on AUROC and AUPRC by 5.7\% and 4.6\%, respectively.


The experimental results demonstrate that SiamDTI can extract effective representations for novel drugs and targets, resulting in its superior performance within this scenario and validating its the stronger generalization ability.
\begin{table*}[!htbp]
    \caption{\textbf{Ablation results} in AUROC on there datasets within different scenarios}
    \centering
    \begin{tabular}{cccccc}
        \hline
        \multirow{2}{*}{Scene} & \multicolumn{2}{c}{Ablation Settings} & \multirow{2}{*}{BindingDB} & \multirow{2}{*}{BioSNAP} & \multirow{2}{*}{human}    \\
                  & BAN    & Siam                                                                                                \\
        \hline
        \multirow{3}*{\thead{Known drugs \\ and targets}}
                  & \ding{51} & \ding{55}                   & $0.9625\pm{0.001}$     & $ 0.9055\pm{0.003}$   & $0.9820\pm{0.002}$  \\
                & \ding{55} & \ding{51}                   & $0.9591\pm{0.001}$     & $ 0.9058\pm{0.003}$   & $0.9826\pm{0.001}$    \\
                  & \ding{51} & \ding{51}                   & $0.9625\pm{0.002}$     & $ 0.9085\pm{0.004}$   & $0.9841\pm{0.003}$  \\
        \hline
        \multirow{3}*{\thead{Novel drugs \\ and targets}}
                  & \ding{51} & \ding{55}                   & $0.5545\pm{0.027}$     & $ 0.6158\pm{0.010}$    & $0.8334\pm{0.020}$  \\
                  & \ding{55} & \ding{51}                   & $0.5586\pm{0.019}$     & $ 0.6068\pm{0.014}$    & $0.8830\pm{0.014}$  \\
                  & \ding{51} & \ding{51}                   & $0.5630\pm{0.029}$     & $ 0.6340\pm{0.018}$    & $0.8655\pm{0.010}$  \\
        \hline
    \end{tabular}
    
    \label{tab:AblationTable3}
\end{table*}
\subsection{Ablation Study}
We conduct ablation experiments within two scenarios to evaluate the effectiveness of various SiamDTI components, including employing bilinear attention network to learn pairwise interactions and utilizing siamese network to comprehensively mine protein information. All the models use the same drug and protein coding method during ablation experiments. The experimental results are shown in the Table \ref{tab:AblationTable3}. We used linear splicing instead of the BAN during feature fusion to verify the network's effectiveness. In this case, the model degenerates to learn the similarities between drugs and proteins. The decoding fusion feature is $f=f_d-f_t$, where $f_d$ and $f_t$ represent the pooled feature vectors of the drug and protein feature maps, respectively.Table \ref{tab:AblationTable3} displays the BAN's effectiveness in learning pairwise interaction features.

To verify the siamese network's effectiveness, we only use the DTI channel for prediction, and the feature fusion module uses the BAN. We observe that the model does not perform satisfactorily on novel drugs and targets without the dual channels for information fusion, indicating that the model's generalization performance is inadequate. The experimental results show that combining SiamDTI's dual-channel strategy with the bilinear attention network can further mine useful protein information, improving the prediction capacity of drug-target protein interactions, especially in zero-shot scenarios. 
\begin{figure}[H]
    \centering
    \includegraphics[width=1.0\linewidth]{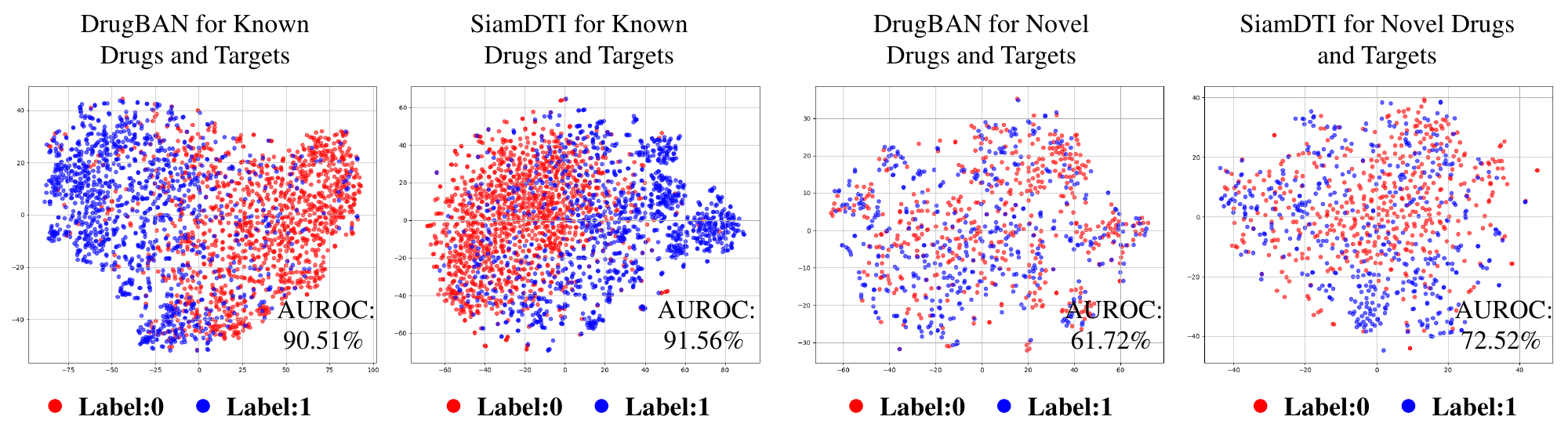}
    \caption{The sample distribution for SiamDTI and Drugban within two scenarios.}
    \label{fig:fig5}
\end{figure}
\subsection{Discussion}
We conduct a theoretical analysis to explore SiamDTI's effectiveness. For SiamDTI, two channels map the input sample to the feature space via a subnetwork of different parameters:
$f_{dt}=F1({p_i,g_j}|p_i \in P, g_j \in G;\theta_1)$ and
$f_{pp}=F2({g_j,g_j}| g_j \in G;\theta_2)$.
The \( f_{dt} \) and \( f_{pp} \) represent the DTI features and the PPI features, respectively. \( \theta_1 \) and \( \theta_2 \) are the two channels' parameters.
According to equation \ref{con:F}–\ref{con:Loss}, the loss function can be expressed as:
\begin{equation}
L_1=-\mathrm{log}(p_1)=-\mathrm{log}(\frac{1}{1+e^{-(w_0f_1+b_0)}})=\mathrm{log}(1+e^{-(w_0f_1+b_0)}),
\end{equation}
\begin{equation}
L_0=-\mathrm{log}(1-p_0)=-\mathrm{log}(1-\frac{1}{1+e^{-(w_0f_0+b_0)}})=\mathrm{log}(1+e^{(w_0f_0+b_0)}),
\end{equation}
\(L_1\) and \(f_1\) represent the loss function and representation for the sample labeled 1, while \(L_0\) and \(f_0\) denote that of the sample labeled 0, respectively.
To minimize the loss function, the formula $w_0f_1+b_0\gg w_0f_0+b_0$ must be satisfied. Thus, it can be expressed as:
\begin{equation}
w_0(f_1-f_0)=\left \| w_0 \right \| \left \| f_1-f_0 \right \| \cos \alpha\gg 0,
\end{equation}
where \(\alpha\) denotes the angle between the vector \(w_0\) and \((f_1 - f_0)\). The above formula demonstrates that SiamDTI's network architecture design and loss function can increase \(\left \| f_1 - f_0 \right \|\), thereby expanding the distance between samples with different labels. Similarly, SiamDTI can reduce the distance between samples with the same label.
Figure \ref{fig:fig5} shows the sample distribution corresponding to the two different methods.
We observe that, within both scenarios, SiamDTI's sample distribution with identical label is more concentrated than that of DrugBAN.
\label{discussion} 

\section{Conclusion}
In this paper, we propose the SiamDTI method, which is a significant advancement in DTI prediction. The SiamDTI model leverages cross-field supervised learning for feature enhancement to obtain more useful target protein information. Its innovative dual-channel framework provides a comprehensive DTI prediction method that improves result accuracy and robustness.
The extensive experimental results demonstrate that SiamDTI outperforms existing mainstream methods during DTI tasks, especially in scenarios involving novel drugs and targets that are more aligned with practical drug development needs. 
This work demonstrates the value of the previously unexplored dual-channel information fusion strategy during DTI prediction tasks by incorporating the PPI channel to integrate more useful target protein information.
Our method is expected to transition the DTI prediction task from the laboratory stage to the large-scale drug screening phase for pharmaceutical companies. 
\newpage

\bibliographystyle{unsrt}
\bibliography{neurips_2024}
\newpage
\appendix
\section{Appendix}

\subsection{Metrics.}
We use AUROC, AUPRC, and F1 score as evaluation metrics to assess the performance of the model in DTI tasks. We conducte five independent rounds of experiments for different testing scenarios using different random seeds. Additionally, We report the mean and standard deviation of each metric.
\label{app:Metrics}
\subsection{Datasets}
To evaluate the performance of the proposed model in DTI tasks, we conducte extensive experiments on three datasets: BindingDB\cite{gao2018interpretable}, BioSNAP\cite{wishart2008drugbank}, and human\cite{liu2015improving}. To investigate the effectiveness and generalization of the model in complex scenarios, we adopt different methods of partitioning datasets for two different application scenarios. For known drugs and targets, there is overlap between the drugs and protein types in the training and test sets. On the other hand, for novel drugs and targets, the types of drugs and targets in the training and test sets do not overlap. Hence, the drug-target pairs in test datasets are entirely novel to the model.
BindingDB is the first publicly available molecular recognition dataset, applied in drug discovery, pharmacology, and related fields. The BindingDB dataset is a custom subset of the Binding dataset\cite{gilson2016bindingdb}, containing affinity information between similar small molecules and target proteins. The BindingDB dataset consists of 39,747 positive samples and 31,218 negative samples.
BioSNAP dataset originates from the DrugBank database, including 4,510 drugs and 2,181 proteins. BioSNAP is a balanced dataset containing effective positive samples and an equal number of randomly selected drug-target pairs as negative samples.
The human dataset consists of high-confidence positive compound–protein interaction (CPI) and negative CPI samples extracted by a filtering framework based on similarity rules. We construct a balanced human dataset using an equal number of positive and negative samples.
\label{app:Datasets}

\subsection{Limitations} 
Although SiamDTI method provides important help for drug research and development, it still has some limitations in practical application.

SiamDTI is a kind of representation learning method. It usually requires a large amount of data to train and optimize the model to obtain accurate predictions. However, in the field of drug discovery, the available data on drug-target interactions are often very limited, which can lead to poorly trained models and inaccurate predictions.

The lack of wet experiments (i.e. actual biochemical or cellular experiments) makes it impossible to truly verify the usefulness of the model.


\label{app:Limitations}

\newpage

\section*{NeurIPS Paper Checklist}

\begin{enumerate}

\item {\bf Claims}
    \item[] Question: Do the main claims made in the abstract and introduction accurately reflect the paper's contributions and scope?
    \item[] Answer: \answerYes{} 
    \item[] Justification: The abstract and introduction of this paper reflect the main contributions of this work. We discuss the motivation of this research work, and analyze the problems existing in the existing methods in the abstract and introduction. On this basis, we propose the main contributions of this work. Then, the paper describes the specific process of the proposed method. Finally, the effectiveness of the proposed method is verified by theoretical analysis and experimental results.

\item {\bf Limitations}
    \item[] Question: Does the paper discuss the limitations of the work performed by the authors?
    \item[] Answer: \answerYes{} 
    \item[] Justification: A discussion of the limitations of this article is in the Appendix \ref{app:Limitations} 

\item {\bf Theory Assumptions and Proofs}
    \item[] Question: For each theoretical result, does the paper provide the full set of assumptions and a complete (and correct) proof?
    \item[] Answer: \answerYes{} 
    \item[] Justification: In Section \ref{discussion}, we analyze the theoretical feasibility of the proposed method and visually verify the effectiveness of the proposed method.

    \item {\bf Experimental Result Reproducibility}
    \item[] Question: Does the paper fully disclose all the information needed to reproduce the main experimental results of the paper to the extent that it affects the main claims and/or conclusions of the paper (regardless of whether the code and data are provided or not)?
    \item[] Answer:\answerYes{} 
    \item[] Justification: We come up with a new model and make the code and data public so that others could access the model.

\item {\bf Open access to data and code}
    \item[] Question: Does the paper provide open access to the data and code, with sufficient instructions to faithfully reproduce the main experimental results, as described in supplemental material?
    \item[] Answer: \answerYes{} 
    \item[] Justification: The code and data are available at:\href{https://anonymous.4open.science/r/DDDTI-434D/}{https://anonymous.4open.science/r/DDDTI-434D/}. We provide the environment requirements as well as instructions and commands for running the code in this link. The data we provide is raw and can be used without pre-processing. Additionally, we provided a script to reproduce all the experimental results.

\item {\bf Experimental Setting/Details}
    \item[] Question: Does the paper specify all the training and test details (e.g., data splits, hyperparameters, how they were chosen, type of optimizer, etc.) necessary to understand the results?
    \item[] Answer: \answerYes{} 
    \item[] Justification: We present the full experimental details in Section \ref{experiment}. It includes the setting of hyperparameters and the introduction of comparison methods and so on.

\item {\bf Experiment Statistical Significance}
    \item[] Question: Does the paper report error bars suitably and correctly defined or other appropriate information about the statistical significance of the experiments?
    \item[] Answer: \answerYes{} 
    \item[] Justification: We performed performance comparisons on three datasets. The experimental results in this paper are taken from the average of five randomized trials, and the size of standard deviation is provided. Some results are accompanied by error bars of standard deviation. We can see it in Figure \ref{fig:fig3}, Table \ref{tab:table1}, and Table \ref{tab:AblationTable3}.

\item {\bf Experiments Compute Resources}
    \item[] Question: For each experiment, does the paper provide sufficient information on the computer resources (type of compute workers, memory, time of execution) needed to reproduce the experiments?
    \item[] Answer: \answerYes{} 
    \item[] Justification: In Section \ref{Complexity Analysis}, we analyze the complexity of the model and indicate the types of cpus and Gpus used in this paper. In addition, we also provide model complexity and model parameter number analysis
    
\item {\bf Code Of Ethics}
    \item[] Question: Does the research conducted in the paper conform, in every respect, with the NeurIPS Code of Ethics \url{https://neurips.cc/public/EthicsGuidelines}?
    \item[] Answer: \answerYes{} 
    \item[] Justification: We have read the NeurIPS Code of Ethics, and the research in this paper complies with the NeurIPS Code of Ethics in every respect

\item {\bf Broader Impacts}
    \item[] Question: Does the paper discuss both potential positive societal impacts and negative societal impacts of the work performed?
    \item[] Answer:\answerYes{}  
    \item[] Justification: We discuss the positive impact of this research on society in the introduction and conclusion.
    
\item {\bf Safeguards}
    \item[] Question: Does the paper describe safeguards that have been put in place for responsible release of data or models that have a high risk for misuse (e.g., pretrained language models, image generators, or scraped datasets)?
    \item[] Answer: \answerNA{} 
    \item[] Justification: This paper poses no such risks.

\item {\bf Licenses for existing assets}
    \item[] Question: Are the creators or original owners of assets (e.g., code, data, models), used in the paper, properly credited and are the license and terms of use explicitly mentioned and properly respected?
    \item[] Answer:\answerYes{} 
    \item[] Justification:We cite all the original papers with comparison methods and datasets.

\item {\bf New Assets}
    \item[] Question: Are new assets introduced in the paper well documented and is the documentation provided alongside the assets?
    \item[] Answer:  \answerNA{} 
    \item[] Justification: The paper does not release new assets.

\item {\bf Crowdsourcing and Research with Human Subjects}
    \item[] Question: For crowdsourcing experiments and research with human subjects, does the paper include the full text of instructions given to participants and screenshots, if applicable, as well as details about compensation (if any)? 
    \item[] Answer: \answerNA{} 
    \item[] Justification: The paper does not involve crowdsourcing nor research with human subjects.

\item {\bf Institutional Review Board (IRB) Approvals or Equivalent for Research with Human Subjects}
    \item[] Question: Does the paper describe potential risks incurred by study participants, whether such risks were disclosed to the subjects, and whether Institutional Review Board (IRB) approvals (or an equivalent approval/review based on the requirements of your country or institution) were obtained?
    \item[] Answer: \answerNA{} 
    \item[] Justification: The paper does not involve crowdsourcing nor research with human subjects.

\end{enumerate}

\end{document}